# Generation of Automatic and Realistic Artificial Profiles

Abigail Paradise, Dvir Cohen, Asaf Shabtai, Rami Puzis
*Department of Software and Information Systems Engineering*
*Ben-Gurion University of the Negev*
Beer-Sheva, Israel.
{abigailp, dvircohe, shabtaia, puzis}@bgu.ac.il

*Abstract*—Online social networks (OSNs) are abused by cyber criminals for various malicious activities. One of the most effective approaches for detecting malicious activity in OSNs involves the use of social network honeypots – artificial profiles that are deliberately planted within OSNs in order to attract abusers. Honeypot profiles have been used in detecting spammers, potential cyber attackers, and advanced attackers. Therefore, there is a growing need for the ability to reliably generate realistic artificial honeypot profiles in OSNs. In this research we present 'ProfileGen' - a method for the automated generation of profiles for professional social networks, giving particular attention to producing realistic education and employment records. 'ProfileGen' creates honeypot profiles that are similar to actual data by extrapolating the characteristics and properties of real data items. Evaluation by 70 domain experts confirms the method's ability to generate realistic artificial profiles that are indistinguishable from real profiles, demonstrating that our method can be applied to generate realistic artificial profiles for a wide range of applications.

## I. Introduction

Online social networks (OSNs) are abused by cyber criminals who exploit the platform to achieve various malicious goals – spreading dubious rumors, misinformation, or propaganda [1], spam and malware distribution [2], harvesting personal information about individuals [3], infiltration of organizations [4-6], and performing industrial espionage [7].

For example, professional social networking platforms can be used to collect intelligence about target organizations during the reconnaissance stage of advanced persistent threats (APTs) [8-10]. The information collected is used to select employees that can be exploited in order to penetrate the organization using social engineering methods (an email message with a malicious URL or payload). Detecting such reconnaissance activity is extremely hard, because it is performed outside the organization's premises.

One of the most effective approaches for detecting malicious activity in OSNs involves the use of social network honeypots – artificial profiles that are deliberately planted within OSNs in order to attract abusers. Honeypot profiles have been successfully used to detect spammers [11-14].

Given their effectiveness and widespread use, there is a growing need for automated frameworks for the creation and management of social network honeypot accounts. Paradise et al. [15] proposed such a framework to aid organizations in the detection of APTs. While their framework supported the automatic generation of basic profile information, such as name, email, address, etc., and automatic wiring (automatically deploying the honeypots) of the honeypots within OSNs, it doesn't eliminate the need for an HR (human resources) expert to create reliable professional experience (such as education and employment) data for the honeypot profiles. Unreliable professional records allow attackers to easily identify honeypots as easily as they identify simple socialbots.

Today, professional profile information is readily available and accessible on the Web on professional social media sites such as LinkedIn. The information copied from existing profiles can be used to populate artificial profiles, however the use of such information results in ethical problems. Such use can violate user privacy. Privacy is considered violated if information intended for a particular audience becomes available to another audience [16], allowing the profile information to be used for other purpose and audience. Additionally, using existing profiles can expose the users to several threats, e.g., it is easier for attackers to identify the artificial profile, the user profile can be de-anonymized (revealing the identity of the person) using the person's posts on OSNs [17] or their group memberships [18], etc. It has also been shown that the social network accounts of users can be identified, even if they have taken precautions against an attack by anonymizing their profiles on OSNs [17]. Furthermore, access to such information can also result in identity theft [19].

In order to create large numbers of realistic social network honeypots without compromising users' privacy, we propose 'ProfileGen' – a novel method for the automated generation of profiles that seamlessly match the position and role that the honeypot pretends to hold, giving particular attention to producing realistic education and employment records. No human intervention is required. The method creates a Markov model from real data, which generates artificial profiles that are indistinguishable from real profiles. Unlike records of personal information, the Markov model cannot be traced back to the OSN profiles from which it was built and thus can be stored offline without violating the OSN user license agreements.

A high quality honeypot is defined as an artificial data item that is similar to real tokens, such that even an expert in the relevant domain will not be able to distinguish the honeypot from real tokens [19]. We performed a field study with experts in the human resources domain in order to evaluate the method's ability to generate realistic artificial profiles that experts cannot distinguish from real profiles. Results confirm the high quality of the generated profiles.

## II. SYNTHETIC DATA GENERATION

Synthetic data generation plays an important role in many different fields, including: prediction of users' movement [21-23] and system evaluation and testing [24]. In order to create realistic artificial profiles, there is also a need to generate synthetic data.

Several studies have focused on relational database generation [25-27], creating artificial database records [20], the generation of digital honeytokens consisting of relational information such as names and addresses [28], and generating basic data for the creation of honeypot profiles in OSNs [11-14]. In the abovementioned studies [11-14, 25-28] the synthetic information generated was basic relational data, but not the more complicated sequential data. Our study includes the generation of sequential data with particular attention to producing realistic education and employment records. To the best of our knowledge, our method is the first generic method for generating artificial profiles with realistic sequential data, such as education and employment records.

Other research has centered the generation of: spatiotemporal datasets [21-23], a set of moving points [21], synthetic trajectories for the positioning data of cellular devices [22], sequences of locations and times by sampling the spatial and temporal probability distributions from call detail records [23], and synthetic meteorological data [29,31].

The Markov chain modeling approach [33] has frequently been used for generating synthetic sequential data, including: daily maximum and minimum temperatures [29], annual rainfall data [30], and wind speed data [31]. Our method uses the Markov model to generate realistic education and employment records for artificial profiles. The method was described shortly in [32].

## III. PROPOSED METHOD

The proposed method for the automated generation of realistic profiles, 'ProfileGen,' pays particular attention to producing sequential data needed for education and employment records.

The process of generating realistic artificial profiles consists of three main phases which are depicted in Figure 1 and described below. 'ProfileGen' relies on real data from curriculum vitae (CVs) which is provided as input. In order to create artificial profiles that appear real, we automatically apply a Markov model that is extracted from the real data (phase 1), and generate profiles accordingly (phase 2). Finally, we examine the similarity of each artificial profile to the corpus of real profiles and remove the profiles that are less similar to the real ones and easier to identify as artificial (phase 3).

Note that we create the artificial profile based on a Markov model. Profile creation can also be handled by more sophisticated approaches such as recurrent neural networks, however since the Markov model provided satisfactory results in the field study described in Section IV, there was no need to use a more sophisticated method.

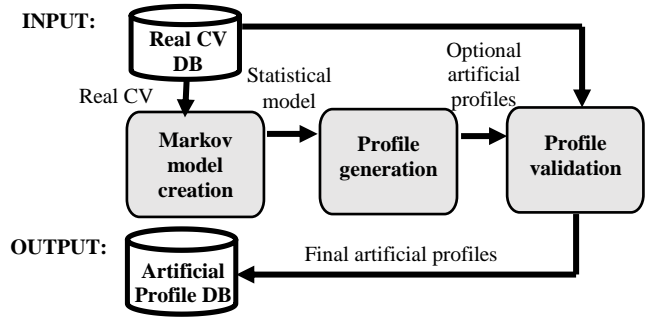

Fig. 1. The proposed method's overall framework.

### A. Markov Model Creation Phase

The input of the proposed method includes a set of real CVs, which can be acquired from an open source database or an HR company. This input forms the real CV database.

*1) Markov model for employment:* Given a set of real CVs, a Markov model is created for the employment records. A Markov model is a random process that represents a system of elements which undergoes transitions from one state to another. We use a Markov model of first order, in which each state depends only on the immediately preceding state.

In order to create a Markov model for the employment records, we refer to each position in the database of real CVs as a state in our model. The Markov model specifies the probability that a user will move from one position to another position and enables the creation of sequences which undergo transitions from one position to another. We use the employment Markov model in the profile generation phase to create the employment record.

*2) Markov model for education:* A Markov model is created for the education records. We refer to each education type in the database of real CVs as a state in our model. The Markov model specifies the probability to move from one education type to another education type and enables the creation of sequences which undergo transitions from one education type to another. We use the education Markov model in the profile generation phase to create the education record. Note that we did not create a single Markov model for both education and employment, because in our preliminary research we found that a single model results in inadequate sequences, since there may be an overlap in the time periods for education and employment. Using two separate Markov models allows for such overlap between employment and education.

### B. Profile Generation Phase

The country of the profile is selected as an input. If a country is not selected, the most common country in the real CV database is selected.

*1) Full name:* A full name is selected from the real CV database by randomly selecting a first and randomly selecting last name based on the country that was selected for the profile.

*2) Employment record*: First, we determine the number of periods of employment in the employment record; the number of periods of employment is defined as the number of positions in the employment record, and the same position can be repeated a number of times in a sequence but with different places of employment. The number of periods of employment is based on the distribution of the number of periods that exist in the real CV database; we choose a number with a probability proportional to its frequency of occurrence in the real CV database (an example of the distribution computed from the dataset we used in this research is presented in Figure 5). Next, we define a sequence of positions for the employment record using the Markov model; each position's transition is defined based on a probability proportional to the position transition frequency of occurrence in the real CV database. Figure 2 presents an example of an artificial sequence of positions created by the Markov model.

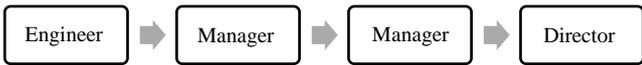

Fig. 2. An example of a position sequence.

The following information is selected for each position in the sequence based on a probability proportional to the frequency of occurrence in the real CV database: a place of employment (with location of employment), duration, and tasks performed at the position. For example, for the position of engineer, a place of employment is defined according to the distribution of places of employment for the engineer position.

In addition, each location of employment is checked in order to verify that it is within the radius of the other employment locations of the profile; if it is not within the radius another place of employment is selected. This verification process helps to avoid a situation in which places of employment are located too far from each other.

Next, we define the starting dates for the employment record. The first period of employment starts at a time that is around the average age to start the first period of employment found in the real CV database. Each new period of employment starts after an average amount of time (a random period of time that is around the average time to begin a new period of employment found in the real CV database).

*3) Education record:* First, we determine the number of education types in the education record (an example of such a sequence is presented in Figure 3). The number of education types is based on the distribution of the number of education types that exist in the real CV database (an example of the distribution computed from the dataset we used in this research is presented in Figure 6), and we choose a number with a probability proportional to its frequency of occurrence in the real CV database. In phase 3 we will check whether the combined education and employment record is realistic.

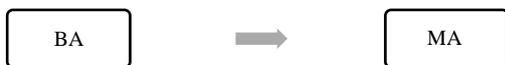

Fig. 3. An example of an education type sequence.

The following information is selected for each item in the education type's sequence: educational institution, field of study, and duration. These were defined based on their distribution, with probabilities proportional to the frequency of occurrence in the real CV database. In addition, each location is checked in order to verify that it is within the radius of the other locations in the education and employment record of the profile; if it is not within this radius another educational institution is selected.

The first period of education starts at a time that is around the average age to start the first period of education found in the real CV database. Each new period of education starts after an average amount of time (a random period of time that is around the average time to begin a new period of education found in the real CV database).

*4) Optional information*: Additional information may be added to the profile, such as: technical and professional qualifications, skills, grants, and awards. This information can also be created using a Markov model which chooses the next item according to its frequency in the real CV database based on the employment and education records.

*5) Specifying current address*: The profile address is based on the location of the last place of employment.

*6) Defining age:* The duration of the items listed in the employment record (in years) is summed; the average age in the real CV database for the first employment period (a random number that is around the average age) is then added to this sum in order to select the age.

*C. Profile Validation Phase*

After creating a database of artificial profiles, we remove artificial profiles with the lowest posterior probability according to the following two heuristics:

*1) Sequence order check:* In this check we look at each position in the employment and education records, in order to determine if the position is relatively close (within a certain threshold) to its average position in the real CV database. For example, intern usually appears at the beginning of employment records in the real database.

Equation (1) checks the sequence order of each item $i$ (position or education type) in an education or employment record. $pos_{sequence}(i)$ is the item's position in the sequence, and $\overline{pos_{database}}$ is the average item position in the real CV database.

A profile for which each position in the employment record and each education type falls within the range of the corresponding average position progresses to the likelihood rank phase – otherwise the profile is rejected.

$$Error(i) = |pos_{sequence}(i) - \overline{pos_{database}}(i)| \qquad (1)$$

*2) Likelihood rank:* We rank and sort the artificial profiles by calculating a likelihood rank that estimates the artificial profile's similarity to the real CVs. A high rank indicates that the artificial profile has a likelihood of being similar to real CVs. This helps us remove the profiles for which the combined education and employment records were not realistic enough.

In order to rank the profiles, for each profile we combine the education type sequence with the employment sequence into one sequence that contains them both, the sequence order was listed in chronological order based on the starting date. Figure 4 combines the sequences presented in Figures 2 and 3.

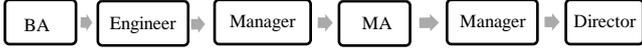

Fig. 4. An example of a combined sequence of education types and positions.

We use Equation (2) to calculate the likelihood rank [34]. Profiles with a likelihood rank lower than a specified threshold are removed. In the field study described in Section IV we removed the profiles that received a likelihood rank of 0. $P_1 P_2 \ldots P_s$ are the profile's positions and education types.

$$P(P_1 P_2 \ldots P_s) = \frac{P(P_1 P_2 P_3) * P(P_2 P_3 P_4) * \ldots * P(P_{s-2} P_{s-1} P_s)}{P(P_2 P_3) * P(P_3 P_4) * \ldots P(P_{s-2} P_{s-1})} \quad (2)$$

Since there are cases where the triple $P_1 P_2 P_3$ does not exist in the real CV database, the probability of a triple is calculated by Equation (3):

$$P(P_1 P_2 P_3) = \frac{P(P_1 P_2) * P(P_2 P_3)}{P(P_2)} \quad (3)$$

## IV. PROFILE QUALITY EVALUATION

A high quality artificial profile is one that even an expert (in the relevant domain) will be unable to distinguish from a real profile. Therefore, a Turing-like test was performed to evaluate the quality of the artificial profiles.

### A. Real CV Database Description

In order to evaluate the proposed method for profile generation, we used the dataset from a case study that was collected and used in [15]. The dataset contains 20,673 users with details including: first name, last name, age, address, birthdate, education record (educational institutions, years of study, and education types), and employment record (places of employment, years, and positions).

In our study the dataset was used as our real CV database. After the real CV database was created, data preprocessing was performed in order to remove incorrect or incomplete data. This focuses on removing profiles that are missing essential data (e.g., positions) and removing profiles with incorrect data (e.g., profiles with employment duration that does not match the person's age - a 30-year-old with 35 years of employment).

The data contains 380 unique education types and 756 unique positions. The average number of periods of employment in the real CV database is 3.51, and the average number of education types in the education record in the real CV database is 1.62. The average age for the first job in the real CV database is 23.18.

### B. Comparison between Artificial Profiles and the Real CV Database

In this section we compare artificial profiles with the data in the real CV database. We generated 10,000 artificial profiles using 'ProfileGen' and compare these against the real CV database.

*1) Age distribution:* We compare the age distribution of the real CV database and the artificial profiles. Table 1 shows the age statistics (the average age and standard deviation). A t-test showed that there was no statistically significant difference between the results.

TABLE I. AGE STATISTICS

| Measures | Artificial profiles | Real CV database |
|---|---|---|
| Average | 33.065 | 33.842 |
| Standard deviation | 8.981 | 7.378 |

*2) Number of periods of employment and education types:* We compare the distribution of the number of periods of employment and the number of education types of real and artificial profiles. Figure 5 presents the distribution of the number of periods of employment (e.g., 1 on the x-axis represents the percentage of profiles that contain only one period of employment in the employment records). We can see that the distribution is very similar when comparing the real and artificial profiles.

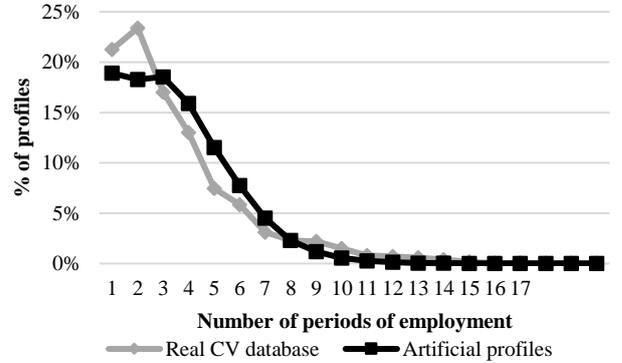

Fig. 5. Distribution of the number of periods of employment.

Figure 6 presents the distribution of the number of periods of education (e.g., 1 on the x-axis represents the percentage of profiles with only one type of education in the education record). The distribution of education periods is also very similar between artificial profiles and the real CV dataset.

Figure 7 presents the number of periods of employment combined with the number of types of education. It can be seen that for both the real CV dataset and artificial profiles most of profiles have between one to six periods.

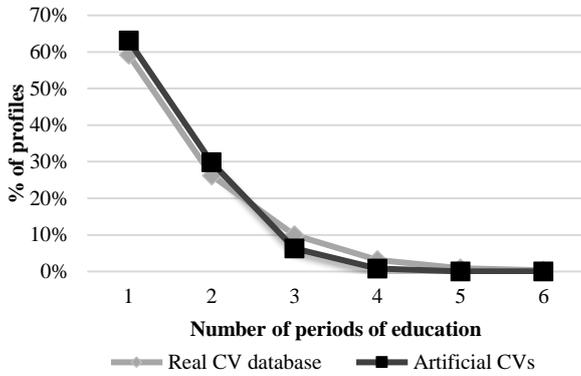

Fig. 6. Distribution of the number of periods of education.

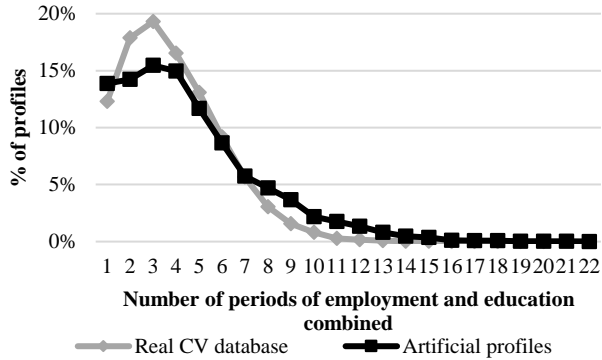

Fig. 7. Distribution of the periods of employment and education combined.

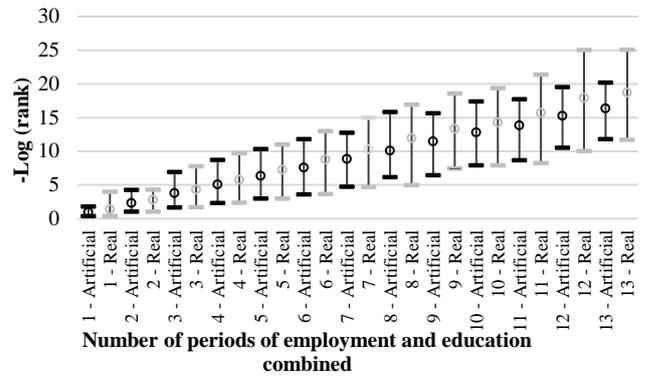

Fig. 8. Average, minimum, and maximum of likelihood rank as a function of the number of periods of employment and education combined.

*3) Likelihood rank:* We compare the likelihood rank of the real CV dataset and artificial profiles. After calculating the likelihood rank of the 10,000 profiles, we found that 2.87% of them received a rank of 0. Those profiles were found to be the least similar to the real CV database; therefore, we defined the threshold as 0 and remove profiles with a rank equal 0.

Figure 8 presents the average, minimum, and maximum likelihood rank of the profiles in the real CV database and the 10,000 artificial profiles (after removing profiles with a rank of 0) as a function of the combined number of periods of employment and education. The likelihood rank is presented in minus log, since the rankings were very low. The lower the rank is, the more realistic the profile is. It can be seen that the rankings of real and artificial profiles are very similar, and this indicates that the artificial profiles are very similar to those in the real CV database. Generally, the rankings of the artificial profiles are lower than the real CV database, and this can be explained by the fact that our method for generating artificial profiles performs optimization of the real CV database, and therefore, on average, the ranks of the real CV database are slightly higher. An additional observation that can be made from Figure 8 is that the likelihood rank increases as the number of periods of employment and education of the profile increases. This is consistent with the distribution of the profile lengths presented in Figure 7, because the longer the combined educational and employment records are, the lower the frequency in the real CV database; therefore profiles with short combined periods of employment and education have the lowest likelihood rank and are preferred.

*4) Profile diversity:* We compare the diversity of artificial profiles against the real CV database. In order to assess the diversity we created a binary matrix of profiles and education types, as well as a binary matrix of profiles and positions for the real CV database and followed the same process for the artificial profiles. We took the position and education types from the real CV database in both cases (for the real CV database and the artificial profiles). We applied t-NSE (an algorithm for dimensionality reduction) on each matrix and then used k-means for creating clusters of the data. We choose the number of clusters based on the Silhouette cofficient; values closer to 1 indicate that the objects are well matched to their clusters.

Figure 9 presents the results of the employment clustering; 80 clusters are sufficient for dividing the data in both the real CV database and the artificial profiles. When comparing the results, both groups received the highest Silhouette cofficient in 80 clusters (0.363 for artificial data and 0.399 for the real CV database), meaning that the clusters in both groups are similarly varied. Figure 10 presents the education clustering. Five clusters received the highest Silhouette score (0.628 for the artificial data and 0.603 for the real CV database). We can see one large cluster that includes Diploma or BA degree and multiple smaller clusters such as BA and MA in both the real CV database and the artificial profiles.

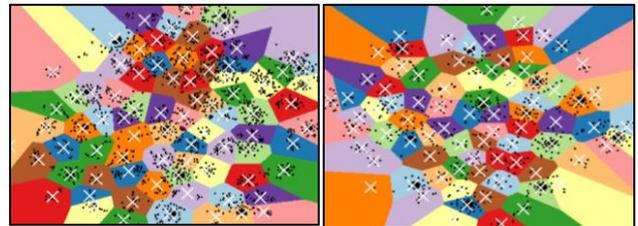

Fig. 9. Employment clustering (right: real CV database, left: artificial profiles).

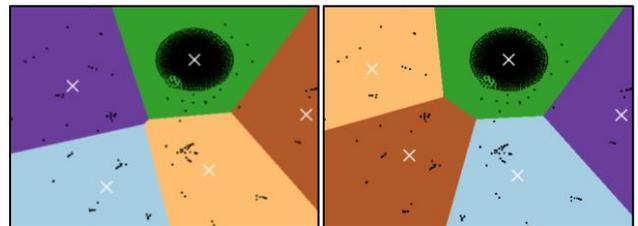

Fig. 10. Education clustering (right: real CV database, left: artificial profiles)

## C. Experimental Setup

We developed a Web-based interactive experiment that included three profile types:

*1) Real profiles*: Randomly selected real profiles of users from the real CV database. After selecting those profiles, we remove them from the real CV database in order to perform the evaluation on separate data; therefore, the real profiles for comparison were deleted from the dataset used for training our Markov model.

*2) Artificial profiles:* Randomly selected profiles created based on our method.

*3) Random profiles:* Randomly selected profiles of users from the real CV database. After selecting those profiles, we changed the following information, replacing the real details with random details: first name, last name, educational record (educational institution, education type), and employment record (place of employment, position).

The main goal of this experiment was to evaluate the quality of the artificial profiles generated. We aimed to make it very difficult (and sometimes close to impossible) for experts to distinguish between real and artificial profiles. Both types of profiles are expected to be significantly more realistic than the random baseline. In order to assess profile reliability we prepared 20 different questionnaires, each of which was completed by 2-7 experts (a total of 70 experts). Each expert was presented with six pairs of profiles (one pair at a time) to make the following comparisons:

- two comparisons between an artificial profile and a real profile,
- two comparisons between an artificial profile and a random profile, and
- two comparisons between a random profile and a real profile.

Each profile contained the following details: full name, country, date of birth, and employment and education records. The order of the comparisons was randomized but the experts were aware that there were both artificial and real profiles. For each pair of profiles the experts were requested to determine whether:

- profile 1 is more realistic than profile 2,
- profile 2 is more realistic than profile 1, or
- both profiles seem equally realistic.

Table 2 summarizes the number of profiles that were selected for the experiment and the number of comparisons made by the experts, using 20 different questionnaires. Note that each profile was compared to other profiles only once and in each questionnaire there were four profiles from each type. Eventually the experiment included a total number of 240 different profiles based on 160 pairwise comparisons.

TABLE II. NUMBER OF PROFILES AND COMPARISONS

| Profile types | Number of profiles | Number of comparisons |
|---|---|---|
| Artificial profiles | 80 | 40 |
| Random profiles | 80 | 40 |
| Real profiles | 80 | 40 |
| Total | 240 | 160 |

Figure 11 presents an example of a comparison between an artificial profile and a real profile. Profile 1 is a random profile, and profile 2 is an artificial profile.

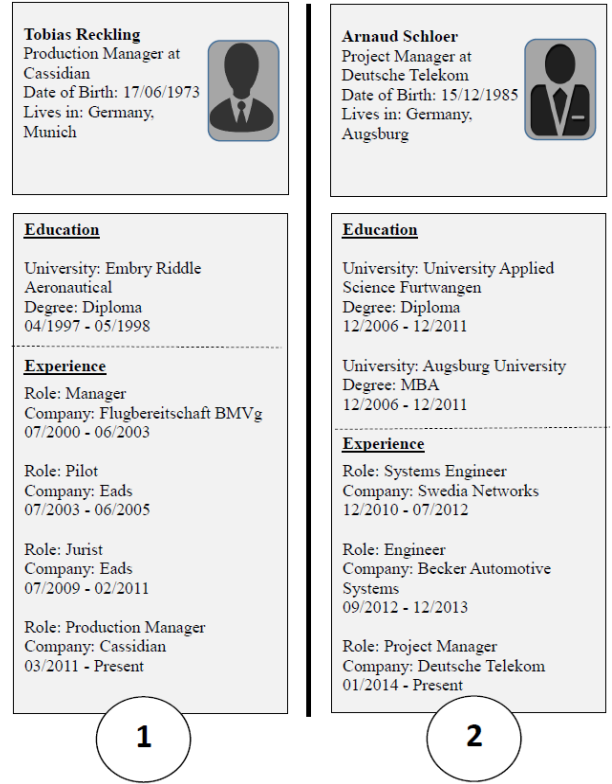

Fig. 11. An example of a comparison between a random profile (1) and an artificial profile (2).

## D. Participants

The 70 experts in the study were experts in the field of human resources, including: HR managers, HR recruiters, and technical recruitment specialists. 69% of the experts were from Israel, 21% from India, 7% from the US, and the remaining were from Germany and England. Our real CV database was mainly based on hi-tech and engineering professions; 92.4% of our experts have experience in hi-tech, engineering, or both.

## E. Results and Discussion

Figure 12 shows the experts' responses in percentages; as expected the real and artificial profiles were considered more realistic than the random profiles. In the comparison between real and artificial profiles, most of the experts found that the profiles were equally realistic.

Note that the experts failed in 12.86% of the cases to select the real profile in the comparison between random and real profiles. We expected a similar error in the comparison between random and artificial profiles to confirm the validity of the proposed approach. As can be seen from the figure, the experts failed in 11.43% of the cases in the comparison between random and artificial profiles, surprisingly, a smaller percentage of errors than that obtained in the comparison between random and real profiles.

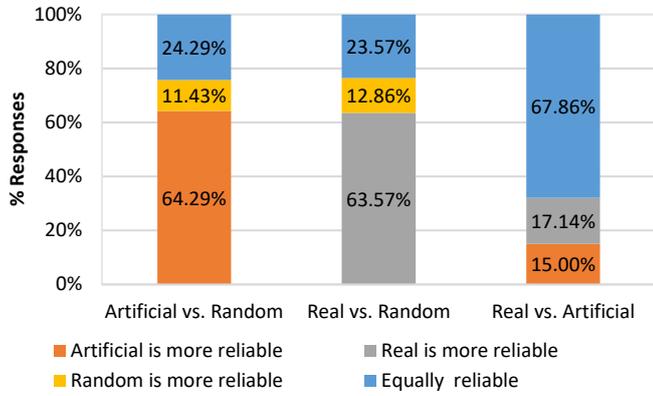

Fig. 12. Experts' responses (in percentages).

To determine whether these results are statistically significant and support our hypotheses we performed two significance tests: a proportion test and a one sample t-test; in addition, we checked for effect size.

*F. Proportion Test*

In order to perform the test, we define the null hypotheses as follows:
- A real profile is as realistic as a random profile, with a probability of 50 (the probability that random selection would be expected to achieve).
- An artificial profile is as realistic as a random profile, with a probability of 50.
- An artificial profile is as realistic as a real profile, with a probability of 50.

The tests that were conducted for real vs. random profiles and for artificial vs. random profiles showed that the confidence interval was 67.37, 82.25 for real vs. random profiles and 68.52, 83.19 for artificial vs. random profiles (p-value<0.05). Because the confidence interval is greater than the value of 0.5 (the random detection rate), we can reject the null hypothesis and accept the alternative hypothesis that real profiles and artificial profiles are more realistic than random profiles.

The test performed for artificial and real profiles showed that the confidence interval was 42.49, 59.61 (p-value>0.05). Because this confidence interval encompasses the value 50, we fail to reject the null hypothesis and would not accept the alternative hypothesis that the profiles are not equally realistic.

*G. T-Test*

In order to conduct the test, we converted the experts' responses (the categories) to numeric values:
- 0 - both profiles seem equally realistic
- 1 - profile 1 is more realistic than profile 2
- -1 - profile 2 is more realistic than profile 1

We define the hypothesis as follows: the mean of experts' responses for artificial vs. random profiles, real vs. random profiles, and real vs. artificial profiles is equal to zero (the profiles are equally realistic).

The t-tests that were conducted for real vs. random profiles and artificial vs. random profiles showed that the differences are considered to be extremely significant with p-value< 0.0001. Therefore, we can reject the null hypothesis and accept the alternative hypothesis that real profiles and artificial profiles are more realistic than random profiles.

The t-test that was conducted for real vs. artificial profiles showed that the difference is not considered statistically significant with p-value = 0.65. Therefore, we fail to reject the null hypothesis and would not accept the alternative hypothesis that the profiles are not equally realistic.

*H. Effect Size*

A significance test does not measure the size of a difference between two groups and a significance test can be misleading when using small sample sizes, [35]. Therefore, we decided to perform the effect size test in addition to the proportion test and t-test described above. Effect size estimates the magnitude of the difference between groups [35]. The standardized mean effect represents the mean difference between two groups in standard deviation units [35-36].

The null hypothesis was defined as follows:

- Real and random profiles are equally realistic.
- Artificial and random profiles are equally realistic.
- Real and artificial profiles are equally realistic.

For real and random profiles (sample size = 140), there was a statistically significant difference between the groups (average = 0.507, standard deviation = 0.712). Furthermore, Cohen's effect size value ($d = 1.08$) suggested high practical significance. Therefore, we can reject the null hypothesis that the real and random profiles are equally realistic.

For artificial vs. random profiles (sample size = 140), there was also a statistically significant difference between the groups (average = 0.528, standard deviation = 0.691). Cohen's effect size value ($d = 1.08$) suggested high practical significance. Therefore, we can reject the null hypothesis that the artificial and random profiles are equally realistic.

For real vs. artificial profiles (sample size = 140), there was no statistically significant difference between real and artificial profiles (average = 0.03, standard deviation = 1). Cohen's effect size value ($d = .04$) suggested low practical significance. Therefore, we fail to reject the null hypothesis and would not accept the alternative hypothesis that the profiles are not equally realistic.

## V. CONCLUSIONS

In this research we propose 'ProfileGen' - a method for the automated generation of realistic profiles. An extensive evaluation by a set of 70 domain experts was performed in order to evaluate our method's ability to generate realistic profiles that cannot be distinguished by experts from real profiles. The results confirm that experts were unable to distinguish between artificial and real profiles, thereby demonstrating that our proposed method can be used to create realistic social network honeypots.

This method is also applicable for generating realistic artificial data for studies that require analysis of employment and education records without exposing the real user profiles. Given the increasing awareness of issues regarding user privacy and strict data protection policies of major social network services, realistic artificial datasets are needed in order to perform academic and commercial research and studies without violating users' privacy.

In future work we plan to develop a user interface to enable to customize the profile generation process. The user interface will enable to edit, remove, and select the type of information that is presented in the profiles.